\documentclass[preprint,12pt]{elsarticle}

\usepackage{amssymb}

\usepackage{amsmath} 
\usepackage{titlesec}   
\usepackage{color}

\journal{Computer Physics Communications}

\begin{document}

\newpagestyle{main}{
    \sethead{}{Computer Physics Communications. DOI: 10.1016/j.cpc.2017.02.006}{}     
    \headrule                                      
}
\pagestyle{main}    

\begin{frontmatter}



\title{A Unified Hamiltonian Solution to Maxwell-Schr\"{o}dinger Equations for Modeling Electromagnetic Field-Particle Interaction}


\author{Yongpin~P.~Chen$^a$}
\author{Wei~E.~I.~Sha$^{b,*}$}
\author{Lijun~Jiang$^{b,*}$}
\author{Min~Meng$^a$}
\author{Yu~Mao~Wu$^c$}
\author{and~Weng~Cho~Chew$^{d,b}$}

\address{$^{a}$ School of Electronic Engineering, University of Electronic Science and Technology of China, Chengdu, 611731, China}

\address{$^{b}$ Department of Electrical and Electronic Engineering, The University of Hong Kong, Hong Kong, China}

\address{$^{c}$ School of Information Science and Technology, Fudan University, Shanghai, 200433, China}

\address{$^{d}$ Department of Electrical and Computer Engineering, University of Illinois at Urbana-Champaign, Urbana, IL 61801 USA}

\cortext[cor1]{wsha@eee.hku.hk(W. E. I. Sha); jianglj@hku.hk(L. Jiang)}

\begin{abstract}
A novel unified Hamiltonian approach is proposed to solve Maxwell-Schr\"{o}dinger equation for modeling the interaction between classical electromagnetic (EM) fields and particles. Based on the Hamiltonian of electromagnetics and quantum mechanics, a unified Maxwell-Schr\"{o}dinger system is derived by the variational principle. The coupled system is well-posed and symplectic, which ensures energy conserving property during the time evolution. However, due to the disparity of wavelengths of EM waves and that of electron waves, a numerical implementation of the finite-difference time-domain (FDTD) method to the multiscale coupled system is extremely challenging. To overcome this difficulty, a reduced eigenmode expansion technique is first applied to represent the wave function of the particle. Then, a set of ordinary differential equations (ODEs) governing the time evolution of the slowly-varying expansion coefficients are derived to replace the original Schr\"{o}dinger equation. Finally, Maxwell's equations represented by the vector potential with a Coulomb gauge, together with the ODEs, are solved self-consistently. {{For numerical examples, the interaction between EM fields and a particle is investigated for both the closed, open and inhomogeneous electromagnetic systems. The proposed approach not only captures the Rabi oscillation phenomenon in the closed cavity but also captures the effects of radiative decay and shift in the open free space. After comparing with the existing theoretical approximate models}}, it is found that the approximate models break down in certain cases where a rigorous self-consistent approach is needed. This work is helpful for the EM simulation of emerging nanodevices or next-generation quantum electrodynamic systems.
\end{abstract}


\end{frontmatter}


\section{Introduction}
\label{}
Computational electromagnetics (CEM) \cite{Chew2001} is nowadays indispensable in the modeling and simulation of electromagnetic (EM) effects (radiation, scattering, or propagation) in various electronic devices. The applications include, but not limited to, antenna design for communication, radar, or biomedical systems, electromagnetic compatibility or interference (EMC/EMI), signal integrity analysis, etc. As the working frequency of modern electronic systems approaches terahertz or even near-infrared, and the physical dimensions of objects scale down to micron or even nanometers, many challenges arise in the development of traditional CEM techonologies. One of the challenges is the multiphysics simulation due to the fact that the real problem is simultaneously governed by several different physical laws. For instance, the performance of a modern electronic device is not solely governed by the circuit or EM physics, but also depends on thermal, mechanical, or even quantum effects \cite{Shao2012, Peng2000, Chen2012, Wang2010}. Among all these effects, quantum effect becomes  increasingly important due to the rapid development of nanotechnologies. At nanoscales, the interaction of EM field and matter is far richer compared to normal radiowave or microwave applications. Moreover, the effect of EM field on a particle (atom, molecule, or quantum dot, etc) can no longer be modeled by the macroscopically-defined polarization and magnetization. Hence, it is imperative to model the quantum effect of particles directly by solving Schr\"{o}dinger equation. Meanwhile, the particles, which generate quantum current sources, also affect the external EM fields governed by Maxwell's equations. The particles absorb and emit electromagnetic waves due to electronic transitions between different energy levels. Based on the above description, the traditional CEM methodology should be modified to integrate the relevant quantum effect into the classical EM simulation. For this purpose, we develop a novel simulation approach for the coupled Maxwell-Schr\"{o}dinger system in the context of EM field-particle interaction where quantum effect plays an important role.

The solution of the coupled Maxwell-Schr\"{o}dinger system has attracted much interest in the last decade. In \cite{Pierantoni2008} and \cite{Pierantoni2009}, the joint simulation of electronic/electromagnetic characterization of emerging nanodevices has been conducted, where a 3-D transmission line matrix scheme combined with 1-D FDTD is proposed to model the interaction of external EM field with the carbon nanotubes. In \cite{Ahmed2010}, a hybrid approach is proposed for the simulation of nano-devices, where the conventional 1-D FDTD and a locally one-dimensional (LOD) FDTD are implemented to solve the Schr\"{o}dinger and Maxwell's equations, respectively.

In both methods, the procedures of coupling EM and quantum mechanical (QM) equations are essentially the same. The electric and magnetic fields ($\mathbf{E}$ and $\mathbf{H}$) are first obtained by solving Maxwell's equations based on initial conditions. The auxiliary vector (magnetic) and scalar (electric) potentials ($\mathbf{A}$ and $\phi$) can then be obtained. The two terms are injected into the Schr\"{o}dinger equation, where the wave function ($\Psi$) is calculated, and the quantum current is derived. This current is then re-injected into the Maxwell's equations as a self-generated source. This self-consistent cycle is repeated until a steady solution is achieved. It is shown lately that by properly choosing the gauge, such cyclic process can be simplified. In the modeling of the interaction between a nanoplate and 2-D EM fields \cite{Ohnuki2013}, it is reported that if a ``length gauge'' is adopted, the EM fields can be directly inserted into the quantum system to improve the computational efficiency. This practice is also implemented to model the interaction of 1-D EM field with electron confined in various 1-D potentials \cite{Takeuchi2014}.

\begin{figure}
\centering
\includegraphics[width=8cm]{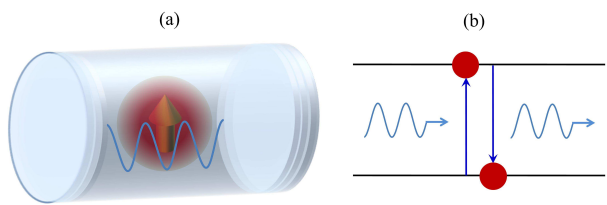}
\caption{(a) A particle is coupled to a resonant cavity. (b) The particle (two-level system) evolves between its ground state and excited state when interacting with the external EM field giving rise to the Rabi oscillation.}\label{FIG1}
\end{figure}

In this work, we will develop a novel unified Hamiltonian approach for numerically solving the Maxwell-Schr\"{o}dinger system. The solution to this system is essential to EM-particle interactions in semi-classical framework. {{Here, the physical model considered is a particle coupled to external EM fields in a closed resonant cavity and in the open free space, which are important to electromagnetic sources (maser, laser, etc) \cite{Oxborrow2012, Ellis2011} and cavity quantum electrodynamics \cite{Andreani1999, Hummer2013}}}. Different from the existing works, a canonical and unified Hamiltonian is developed to derive the Hamilton's equations of the Maxwell-Schr\"{o}dinger system, which ensures the energy conservation of the entire system. For the EM part, the vector potential (not fields) with a Coulomb gauge is updated by the FDTD method that is suitable for arbitrary 3-D dynamic problems. For the QM part, the reduced eigenmode expansion is adopted instead of discretizing the time-dependent Schr\"{o}dinger equation for alleviating the multiscale difficulty. The well-known Rabi oscillation for characterizing (stimulated) emission and absorption processes of the particle is calculated, where the schematic configuration and the process of Rabi oscillation are shown in Fig. \ref{FIG1}. {{Moreover, the effects of radiative decay and radiative shift for a particle in free space are also studied.}} The numerical results are compared with the approximate analytic models \cite{Gerry2005}. It is indicated that in some cases, the analytic model breaks down and a rigorous self-consistent simulation is needed.

The rest of the paper is organized as follows: in Section 2, a unified Hamiltonian is derived for the Maxwell-Schr\"{o}dinger system. A self-consistent approach with the reduced eigenmode expansion is developed in Section 3. After that, in Section 4, several numerical results from the proposed approach are presented and compared with the approximate theoretical models. Particularly, different aspects (including field intensity, cavity loss, etc) affecting the Rabi oscillation are discussed in details. Finally, a brief conclusion is included and possible future works are suggested in Section 5.

\section{Hamilton's Equations of the Electromagnetic and Quantum System}
For the sake of completeness, some basic equations are reviewed first. The curl-form Maxwell's equations are given by \cite{Kong1975}
\begin{equation} \label{EQ1}
\nabla\times \mathbf{E}(\mathbf{r},t) = - \frac{\partial \mathbf{B}(\mathbf{r},t)}{\partial t}
\end{equation}
\begin{equation} \label{EQ2}
\nabla\times \mathbf{H}(\mathbf{r},t) = \frac{\partial \mathbf{D}(\mathbf{r},t)}{\partial t} + \mathbf{J}(\mathbf{r},t).
\end{equation}
On the other hand, the time-dependent Schr\"{o}dinger equation governing dynamics of a particle is of the form \cite{Miller2008}
\begin{equation} \label{EQ3}
\left[\frac{\hat{\mathbf{p}}^2}{2m} + V(\mathbf{r},t)\right]\Psi(\mathbf{r},t)=i\hbar\frac{\partial\Psi(\mathbf{r},t)}{\partial t}
\end{equation}
where $m$ is the mass of the particle, $V(\mathbf{r},t)$ is the potential energy, $\hbar$ is the reduced Plank's constant, and $\hat{\mathbf{p}}$ is the momentum operator
\begin{equation} \label{EQ4}
\hat{\mathbf{p}}=-i\hbar\nabla.
\end{equation}
Under an EM wave illumination, the equation is modified to
\begin{equation} \label{EQ5}
\begin{array}{rl}
&\left\{\frac{1}{2m}\left[\hat{\mathbf{p}}-q\mathbf{A}(\mathbf{r},t)\right]^2 + q\phi(\mathbf{r},t)+ V(\mathbf{r})\right\}\Psi(\mathbf{r},t)\\
&=i\hbar\frac{\partial\Psi(\mathbf{r},t)}{\partial t}
\end{array}
\end{equation}
where $q$ is the electric charge of the particle, $\mathbf{A}(\mathbf{r},t)$ and $\phi(\mathbf{r},t)$ are the magnetic and electric potentials associated with the EM field. The electrostatic potential energy $V$ (such as nuclear-electron coulombic potential or effective confining potential) is assumed to be independent of time.

To be consistent with quantum theory, the classical EM formulation in terms of $\mathbf{E}$ and $\mathbf{H}$ fields should be replaced with the $\mathbf{\mathbf{A}}$-$\phi$ formulation \cite{Masiello2005}, \cite{Chew2014}. In the EM field-particle interaction problem, the Coulomb gauge (radiation or transverse gauge) within the nonrelativistic limit is adopted, i.e.
\begin{equation}\label{EQ6}
\nabla\cdot\mathbf{A} = 0
\end{equation}
{{The Coulomb gauge splits the EM fields into the transverse (electrodynamic) and longitudinal (electrostatic) parts; and the longitudinal part can be directly inserted into the Schr\"{o}dinger equation as shown in Eq. \eqref{EQ5}. Therefore, Coulomb gauge greatly simplifies the quantum optics problem and is commonly adopted \cite{Kira2014}.}} Next, we define an auxiliary variable
\begin{equation}\label{EQ7}
\mathbf{Y} = -\epsilon_0\mathbf{E}
\end{equation}
After that, the total Hamiltonian of the Maxwell-Schr\"{o}dinger system can be expressed as
\begin{equation}\label{EQ8}
{H}\left( {{\bf{A}},{\bf{Y}}},{\Psi,\Psi^{*}} \right) = {{ H}^{em}}\left( {{\bf{A}},{\bf{Y}}} \right)+{{ H}^q}\left( \Psi, \Psi^{*}, \mathbf{A} \right)
\end{equation}
where
\begin{equation}\label{EQ9}
{ H^{em}}\left( {{\bf{A}},{\bf{Y}}} \right) = \int_\Omega {\left( {\frac{1}{2\epsilon_0}{{\left| {\bf{Y}} \right|}^2} + \frac{1}{{2\mu_0}}{{\left| {\nabla  \times {\bf{A}}} \right|}^2}} \right)} d\bf{r}
\end{equation}
\begin{equation}\label{EQ10}
{{ H}^q}\left( \Psi, \Psi^{*}, \mathbf{A} \right)=\int_\Omega {\left[ {{\Psi ^*}\frac{{\left( {\hat {\bf{p}} - q{\bf{A}}} \right)}^2}{{2m}}\Psi  + {\Psi ^*}V\Psi } \right]} d\bf{r}.
\end{equation}
In the above $\mathbf{A}$ and $\mathbf{Y}$ are real valued while $\Psi$ is complex valued. In Eq. \eqref{EQ8}, $H^{em}$ consists of electric and magnetic energy stored in the EM field, and $H^q$ consists the kinetic and potential energy of the quantum system. By invoking the variational principle, the Hamilton's equations of the EM and QM parts can be derived as
\begin{equation}\label{EQ11}
\frac{{\partial {\bf{A}}}}{{\partial t}} = \frac{{\partial {{ H}}}}{{\partial {{\bf{Y}}_{}}}} = \frac{\bf{Y}}{\epsilon_0}
\end{equation}
\begin{equation}\label{EQ12}
\frac{{\partial {\bf{Y}}}}{{\partial t}} =  - \frac{{\partial {{ H}}}}{{\partial {\bf{A}}}} = -\frac{{\nabla  \times \nabla  \times {\bf{A}}}}{{\mu_0 }} + \bf{J}
\end{equation}
\begin{equation}\label{EQ13}
\frac{{\partial \Psi }}{{\partial t}} = \frac{1}{{i\hbar }}\frac{{\partial {{ H}}}}{{\partial {\Psi ^*}}} = \frac{1}{{i\hbar }}\left[ {\frac{{{{\left( {\hat {\bf{p}} - q{\bf{A}}} \right)}^2}}}{{2m}} + V} \right]\Psi
\end{equation}
\begin{equation}\label{EQ14}
\frac{{\partial {\Psi ^*}}}{{\partial t}} = \frac{{ - 1}}{{i\hbar }}\frac{{\partial {{ H}}}}{{\partial \Psi }} =  - \frac{1}{{i\hbar }}\left[ {\frac{{{{\left( {\hat {\bf{p}} + q{\bf{A}}} \right)}^2}}}{{2m}} + V} \right]{\Psi ^*}
\end{equation}
where $\mathbf{J}$ has the expression of
\begin{equation}\label{EQ15}
{\bf{J}} = \frac{q}{{2m}}\left[ {{\Psi ^*}\left( {\hat {\bf{p}} - q{\bf{A}}} \right)\Psi  + \Psi \left( { - \hat {\bf{p}} - q{\bf{A}}} \right){\Psi ^*}} \right]
\end{equation}
which is essentially the QM current generated from the particle illuminated by the external EM fields. The expression of QM current is the same as that of probability current in the continuity equation \cite{Abers2003}; but it can be derived by the variational principle directly. The generated quantum current will in return perturb and deform the EM field. In matrix notation, the Hamilton's equations can be written as
\begin{equation}\label{EQ16}
\left[ {\begin{array}{*{20}{c}}
0&{ 0}&-1&0\\
0&0&0&-1\\
1&0&0&0\\
0&1&0&0
\end{array}} \right]
\left( {\begin{split}
&{\dot{\bf{A}}}\\
&{\dot{\Psi}}\\
&{\dot{\bf{Y }}}\\
&{\dot{{{\Psi}}}^*}
\end{split}} \right) = \left( {\begin{split}
{\frac{{\partial {{ H}}}}{{\partial {\bf{A}}}}}\\
{\frac{1}{{i\hbar }}\frac{{\partial {{ H}}}}{{\partial  \Psi }}}\\
{\frac{{\partial {{ H}}}}{{\partial {\bf{Y }}}}}\\
{\frac{1}{{i\hbar }}\frac{{\partial {{ H}}}}{{\partial {{\Psi}^*}}}}
\end{split}} \right)\
\end{equation}
The above is not a canonical form of Hamilton's equations. To this end, we apply change of variables and split the wave function into the real and imaginary parts in a normalized way
\begin{equation}\label{EQ17}
 \Psi {\rm{ = }}\frac{1}{\sqrt{2\hbar}}\left( \Psi_r + i\Psi_i \right).
\end{equation}
Taking $\Psi_r$ and $\Psi_i$ as new (real-valued) variables, the total system can be re-derived as
\begin{equation}\label{EQ18}
\left[ {\begin{array}{*{20}{c}}
0&{ 0}&-1&0\\
0&0&0&-1\\
1&0&0&0\\
0&1&0&0
\end{array}} \right]
\left( {\begin{split}
&{\dot{\bf{A}}}\\
&{\dot{\Psi_r}}\\
&{\dot{\bf{Y}}}\\
&{\dot{\Psi_i}}
\end{split}} \right) = \left( {\begin{split}
{\frac{{\partial {{ H}}}}{{\partial {\bf{A}}}}}\\
{\frac{{\partial {{ H}}}}{{\partial {\Psi_r}}}}\\
{\frac{{\partial {{ H}}}}{{\partial {\bf{Y }}}}}\\
{\frac{{\partial {{ H}}}}{{\partial {\Psi_i}}}}
\end{split}} \right).
\end{equation}
If we define ``generalized coordinates and momenta'' as
\begin{equation}\label{EQ19}
\mathbf{q}=\left(\mathbf{A},\Psi_r\right)\quad \mathbf{p}=\left(\mathbf{Y},\Psi_i\right)
\end{equation}
the system is shown to be analogous to the standard Hamilton's equations \cite{Miller2008}
\begin{equation}\label{EQ20}
\frac{\partial \mathbf{p}}{\partial t} = - \frac{\partial H}{\partial \mathbf{q}}
\end{equation}
\begin{equation}\label{EQ21}
\frac{\partial \mathbf{q}}{\partial t} = \frac{\partial H}{\partial \mathbf{p}}.
\end{equation}
Finally we have
\begin{equation}\label{EQ22}
\left( {\begin{split}
&{\dot{\bf{A}}}\\
&{\dot{\Psi_r}}\\
&{\dot{\bf{Y}}}\\
&{\dot{\Psi_i}}
\end{split}} \right) = \overline{\mathbf{M}} \cdot \left( {\begin{split}
{\frac{{\partial {{ H}}}}{{\partial {\bf{A}}}}}\\
{\frac{{\partial {{ H}}}}{{\partial {\Psi_r}}}}\\
{\frac{{\partial {{ H}}}}{{\partial {\bf{Y }}}}}\\
{\frac{{\partial {{ H}}}}{{\partial {\Psi_i}}}}
\end{split}} \right)
\end{equation}
where
\begin{equation}\label{EQ23}
\overline{\mathbf{M}} =
\left[ {\begin{array}{*{20}{c}}
0&0&1&0\\
0&0&0&1\\
-1&0&0&0\\
0&-1&0&0
\end{array}} \right] =
\left[ {\begin{array}{*{20}{c}}
\mathbf{0}&\overline{\mathbf{I}}\\
-\overline{\mathbf{I}}&\mathbf{0}
\end{array}} \right]
\end{equation}

Obviously, this system is well-posed, which means the behavior of solution changes continuously with the initial conditions. Also $\overline{\mathbf{M}}$ satisfies the following relation
\begin{equation}\label{EQ24}
\overline{\mathbf{M}}^T\overline{\mathbf{\Omega}}\overline{\mathbf{M}}=\overline{\mathbf{\Omega}}
\end{equation}
where
\begin{equation}\label{EQ25}
\overline{\mathbf{\Omega}} =
\left[ {\begin{array}{*{20}{c}}
\mathbf{0}&\overline{\mathbf{I}}\\
-\overline{\mathbf{I}}&\mathbf{0}
\end{array}} \right].
\end{equation}
Hence, $\overline{\mathbf{M}}$ is a symplectic matrix; and energy (volume) conservation property can be maintained during the time evolution of the Maxwell-Schr\"{o}dinger system \cite{Feng2010, Ren2012}. In other words, if the initial state in the phase space is
\begin{equation}\label{EQ26}
({\bf{A}}^0,{\Psi_r}^0,{\bf{Y }}^0,{\Psi_i}^0;\,t=t_0)
\end{equation}
and after time $t_1$ it becomes
\begin{equation}\label{EQ27}
({\bf{A}}^1,{\Psi_r}^1,{\bf{Y }}^1,{\Psi_i}^1;\,t=t_1)
\end{equation}
we always have
\begin{equation}\label{EQ28}
d{\bf{A}}^0\wedge d{\Psi_r}^0\wedge d{\bf{Y }}^0\wedge d{\Psi_i}^0=
d{\bf{A}}^1\wedge d{\Psi_r}^1\wedge d{\bf{Y }}^1\wedge d{\Psi_i}^1.
\end{equation}
The above exterior-derivative identity suggests that Hamiltonian flows preserve the phase-space volumes \cite{Feng2010}. Physically, it means that total energy of the Maxwell-Schr\"{o}dinger system is conserved.

\begin{figure}
\centering
\includegraphics[width=10cm]{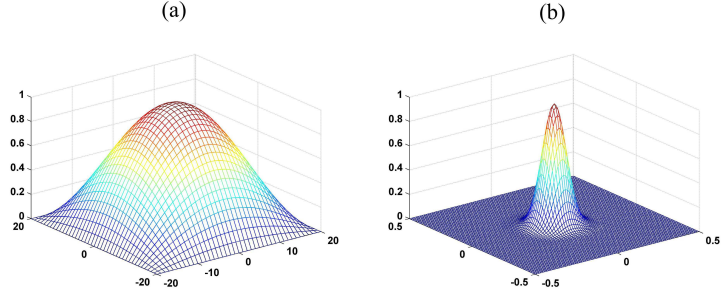}
\caption{The profiles and supports of (a) the EM wave in a cavity, and (b) the particle wave at its ground state.}\label{FIG2}
\end{figure}

In principle, the system in Eq. \eqref{EQ22} can be solved by the conventional FDTD method \cite{Taflove2005}, where all the four variables $\mathbf{A}$, $\Psi_r$, $\mathbf{Y}$, $\Psi_i$ are updated in each time step with initial values and proper boundary conditions. However, the wavelength of the EM wave is much larger than that of the particle wave ($\lambda_{em}\gg\lambda_q $). To show this, the profiles and supports of EM wave in a cavity of $40\times40\times40$ nm$^3$ and a particle wave at its ground state are plotted in Fig. \ref{FIG2}. There is a distinct mismatch of characteristic length between them. In order to guarantee the discretization accuracy, a much refined grid is required around the position of the particle. This multiscale nature leads to a serious efficiency problem in FDTD simulation. In the following section, we will show that this problem can be mitigated by a reduced eigenmode expansion technique.

\section{Self-Consistent Solution with Reduced Eigenmode Expansion}
Without loss of generality, the particle is assumed to be a 3D isotropic quantum harmonic oscillator with two energy levels for ground and excited states. Quantum harmonic oscillator is one of the most important systems in quantum mechanics. It can be applied to model effective confining potential in atoms, molecules and quantum dots \cite{Merkt1991}. The Schr\"{o}dinger equation for the 3D isotropic quantum harmonic oscillator is \cite{Miller2008}
\begin{equation}\label{EQ28_Add}
\left[\frac{\hat{\mathbf{p}}^2}{2m} + \frac{m\omega^2}{2}\mathbf{r}^2\right]\Psi(\mathbf{r},t)=i\hbar\frac{\partial\Psi(\mathbf{r},t)}{\partial t}
\end{equation}
where $\omega$ is the angular frequency of the oscillator. The two corresponding eigenstates (eigenmodes) are denoted as $\psi_g(\mathbf{r})$ and $\psi_e(\mathbf{r})$ with the eigenenergies $E_g=\hbar\omega_g$ and $E_e=\hbar\omega_e$, which are respectively for the ground and excited states. The particle absorbs electromagnetic waves when the electron jumps from ground to excited states. Similarly, the particle emits electromagnetic waves when the electron drops from excited to ground states. According to the reduced eigenmode expansion technique, the time-dependent wave function can be expanded as:
\begin{equation}\label{EQ29}
\Psi(\mathbf{r},t) = a(t)\exp(-i\omega_gt)\psi_g(\mathbf{r}) + b(t)\exp(-i\omega_et)\psi_e(\mathbf{r})
\end{equation}
where $a(t)$ and $b(t)$ are the unknown slowly-varying expansion coefficients. The fast-varying terms $\exp(-i\omega_gt)$ and $\exp(-i\omega_et)$ describe the time evolution of the eigenstates. As shall be seen, only the difference of $\omega_e$ and $\omega_g$, namely the transition frequency $\omega_0=\omega_e-\omega_g$ matters. The squared magnitudes denote the probabilities of occupation of the corresponding quantum states, which satisfy the following probability conserving relation
\begin{equation}\label{eq30}
|a(t)|^2+|b(t)|^2=1.
\end{equation}

By applying the Galerkin test, we have
\begin{equation}\label{eq31}
\left\langle\psi_g\left|\frac{{\partial \Psi }}{{\partial t}}\right.\right\rangle = \left\langle\psi_g\left|\frac{1}{{i\hbar }}\frac{{\partial {{ H}}}}{\partial \Psi^* }\right.\right\rangle
\end{equation}
\begin{equation}\label{eq32}
\left\langle\psi_e\left|\frac{{\partial \Psi }}{{\partial t}}\right.\right\rangle = \left\langle\psi_e\left|\frac{1}{{i\hbar }}\frac{{\partial {{ H}}}}{{\partial \Psi^* }}\right.\right\rangle
\end{equation}
where the inner product is defined as
\begin{equation}\label{EQ33}
\left\langle\psi_i(\mathbf{r})|\psi_j(\mathbf{r})\right\rangle = \int_\Omega d\mathbf{r} \psi_i^*(\mathbf{r})\cdot\psi_j(\mathbf{r}).
\end{equation}
According to orthonormality of the eigenmodes
\begin{equation}\label{EQ34}
\left\langle\psi_i(\mathbf{r})|\psi_j(\mathbf{r})\right\rangle = \left\{ {\begin{split}
&1 \ \ \ \ i=j\\
&0 \ \ \ \ i\neq j
\end{split}} \right.
\end{equation}
and selection rule due to the parity of eigenmodes (the integral of an odd function is equal to zero when it is integrated over the whole of space):
\begin{equation}\label{EQ35}
\left\langle\psi_i(\mathbf{r})|\hat{\mathbf{p}}|\psi_i(\mathbf{r})\right\rangle = 0
\end{equation}
we can arrive at the following two ordinary differential equations:
\begin{equation}\label{EQ36}
i\hbar \frac{d{a(t)}}{dt}=-\frac{q\mathbf{A}}{m}\langle\psi_g|\hat{\mathbf{p}}|\psi_e\rangle b(t) e^{-i\omega_0t} +\frac{q^2\mathbf{A}^2}{2m}a(t)
\end{equation}
\begin{equation}\label{EQ37}
i\hbar\frac{d{b(t)}}{d{t}}=-\frac{q\mathbf{A}}{m}\langle\psi_e|\hat{\mathbf{p}}|\psi_g\rangle a(t) e^{i\omega_0t} +\frac{q^2\mathbf{A}^2}{2m}b(t)
\end{equation}
where the expectation value of quantum current can be expressed as
\begin{equation}\label{EQ38}
\begin{array}{rl}
\langle\mathbf{J}\rangle&=\frac{-q^2}{m} \mathbf{A}\left(|a|^2+|b|^2\right)\\[1em]
&+\frac{q}{m}\left[a^{*}(t)b(t)e^{-i\omega_0t}\langle\psi_g|\hat{\mathbf{p}}|\psi_e\rangle
\right.\\ [1em]
&\left.+b^{*}(t)a(t)e^{i\omega_0t}\langle\psi_e|\hat{\mathbf{p}}|\psi_g\rangle\right]
\end{array}.
\end{equation}
Finally, Eqs. \eqref{EQ11}, \eqref{EQ12}, \eqref{EQ36}, \eqref{EQ37} and \eqref{EQ38} constitute a complete system and can be solved self-consistently by the FDTD method. {{The initial values of the expansion coefficients ($a$ and $b$), and the vector potential $\mathbf{A}$ or the auxiliary variable $\mathbf{Y}$ are needed to be predefined. To guarantee a unique solution, the boundary conditions for the tangential $\mathbf{A}$ should also be specified. For a closed resonant cavity, the tangential $\mathbf{A}$ vanishes at the boundary. For the open free space, however, the convolutional perfectly matched layer (CPML) should be employed to absorb the outgoing waves \cite{Gedney2000}.}}

When the particle (or two-level system) is illuminated by electromagnetic waves in a cavity, it cyclically absorbs photons and re-emits them by (stimulated) emission, which is called Rabi oscillation. If the resonant cavity is ideal without any material loss and radiation (leaky) loss, the emitted electromagnetic waves will react on the particle and thus Rabi oscillation will be cyclic. Once the expansion coefficients in Eq. \eqref{EQ29} are numerically obtained, the population inversion (factor) can be defined as:
\begin{equation} \label{EQ39}
W(t)=|b(t)|^2-|a(t)|^2.
\end{equation}

\section{Numerical Results}
\subsection{Particle in a resonant cavity}
Several numerical results are presented in this section to validate the proposed approach. {{A nanocavity of dimension $L_x=L_y=L_z=40$ nm with a spatial grid of $1$ nm is considered as the resonant cavity.}} For simplicity, the cavity works at the fundamental resonant mode (TE$_{101}$), with an initial condition
\begin{equation} \label{EQ40}
Y_y|_{t=0}=-\epsilon_0A_0\sin\left(\frac{\pi}{L_x}x\right)\sin\left(\frac{\pi}{L_z}z\right)\cos\left(\omega t\right)|_{t=0}
\end{equation}
where
\begin{equation} \label{EQ41}
\omega=c\sqrt{\left(\frac{\pi}{L_x}\right)^2+\left(\frac{\pi}{L_z}\right)^2}.
\end{equation}
A particle resides at the center of this cavity and is initially prepared at a superposition state with $a=1/\sqrt{2}$ and $b=i/\sqrt{2}$. Due to the influence of the external EM fields, the particle oscillates between its ground state and excited state periodically, according to the semi-classical description of the Rabi oscillation \cite{Gerry2005}. This phenomenon can be theoretically predicted by the Rabi model, under certain approximations (see Appendix). We will show in the following that the theoretical approximations can be reproduced by our numerical results in some cases. While in other cases, the approximate model breaks down and a rigorous numerical solution is needed. Besides, by using the proposed numerical methods, the model can be extended to accommodate more complex boundaries and different loss channels.

The FDTD discretizations of Eqs. \eqref{EQ11} and \eqref{EQ12} are briefly summarized for clarity,
\begin{equation}\label{eq42}
\begin{array}{rl}
&A_y^{n+1}(i,j+\frac{1}{2},k) = A_y^n(i,j+\frac{1}{2},k) \\ [1em]
&+ \frac{\Delta t}{\epsilon_0} Y_y^{n+\frac{1}{2}}(i,j+\frac{1}{2},k)
\end{array}
\end{equation}
\begin{equation}\label{eq43}
\begin{array}{rl}
&Y_y^{n+\frac{1}{2}}(i,j+\frac{1}{2},k)=Y_y^{n-\frac{1}{2}}(i,j+\frac{1}{2},k) \\ [1em]
&+\frac{\Delta_t}{\mu_0\Delta_x^2} \left[(A_y^n(i+1,j+\frac{1}{2},k)-2A_y^n(i,j +\frac{1}{2},k) \right. \\ [1em]
&\left.+A_y^n(i-1,j+\frac{1}{2},k)\right] \\ [1em]
&+\frac{\Delta_t}{\mu_0(\Delta_z^2)} \left[(A_y^n(i,j+\frac{1}{2},k+1)-2A_y^n(i,j+\frac{1}{2},k)\right. \\ [1em]
&\left.+A_y^n(i,j+\frac{1}{2},k-1)\right]+J_y^n(i,j+\frac{1}{2},k)
\end{array}
\end{equation}
where $\mathbf{A}$ and $\mathbf{Y}$ are defined on the same collocated grids. The time-stepping schemes of Eqs. \eqref{EQ36} and \eqref{EQ37} are similar and will not be repeated again.


\subsubsection{Effect from Field Intensity}
\begin{figure}
\centering
\includegraphics[width=8cm]{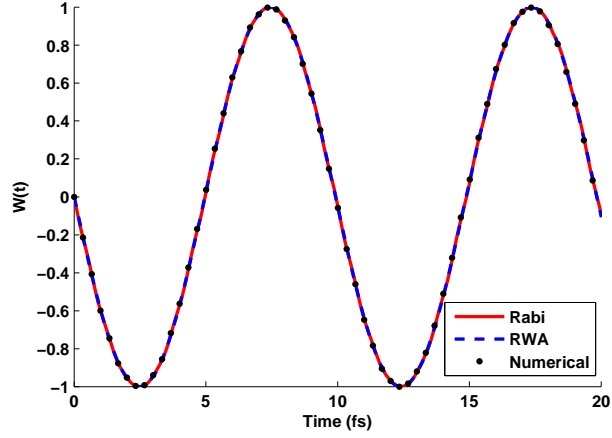}
\caption{Rabi oscillations of population inversion with a weak field ($\Omega=0.02\omega$) in tuning case.}\label{FIG3}
\end{figure}

\begin{figure}
\centering
\includegraphics[width=8cm]{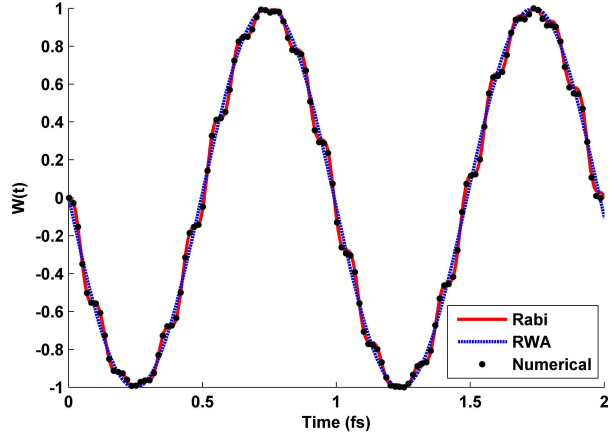}
\caption{Rabi oscillations of population inversion with a strong field ($\Omega=0.2\omega$) in tuning case.}\label{FIG4}
\end{figure}

We first consider the effect from EM field strength. In tuning case ($\Delta=0$), where $\Delta=\omega-\omega_0$, i.e. the resonant working frequency of the cavity $\omega$ is equal to the transition frequency of the particle $\omega_0$ (see Appendix). In this case, the Rabi frequency ($\Omega_R=\Omega$) is proportional to the intensity of $\mathbf{E}$ field [Eqs. \eqref{EQ53}--\eqref{EQ52}]. We compare the weak field case with $\Omega=0.02\omega$ and strong field case with $\Omega=0.2\omega$. Rabi oscillations of population inversion are calculated and shown in Fig. \ref{FIG3} and Fig. \ref{FIG4}. Here, three methods including the theoretical Rabi model [Eqs. \eqref{EQ46}--\eqref{EQ47}], rotating wave approximation (RWA) [Eqs. \eqref{EQ48}--\eqref{EQ49}], and the proposed numerical approach, are adopted. For the weak field case, all the curves agree well with each other. For the strong field case, however, due to the fact that the Rabi frequency is comparable with the EM frequency, the RWA is no longer an appropriate approximation. The fast oscillatory term with a frequency of $\omega+\omega_0$ [Eqs. \eqref{EQ46}--\eqref{EQ47}] is observable. It is shown that our numerical result captures this high frequency modulation well and agrees with the Rabi model.

\subsubsection{Effect from Detuning}
\begin{figure}
\centering
\includegraphics[width=8cm]{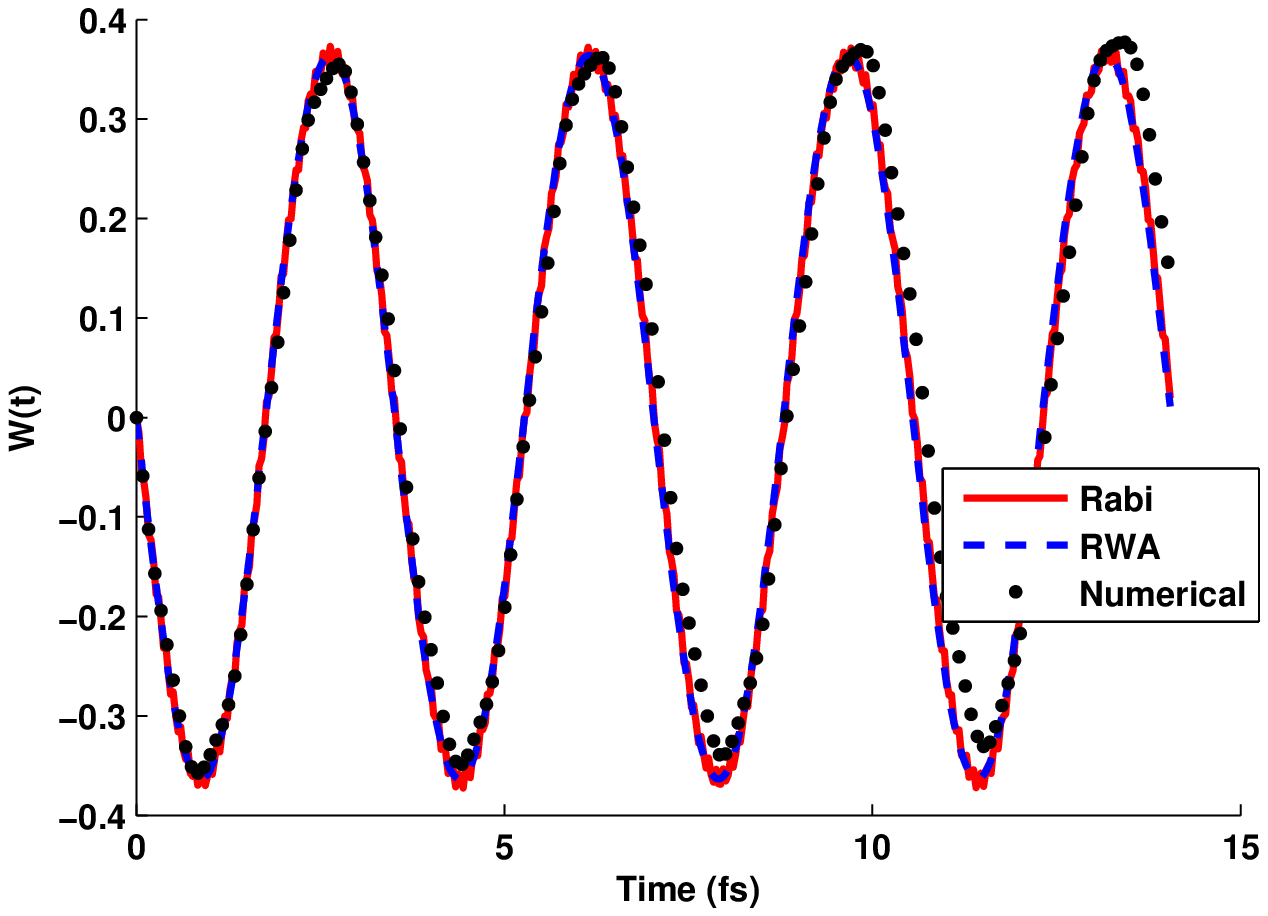}
\caption{Rabi oscillations of population inversion with a small detuning $\Delta=0.05\omega$.}\label{FIG5}
\end{figure}

\begin{figure}
\centering
\includegraphics[width=8cm]{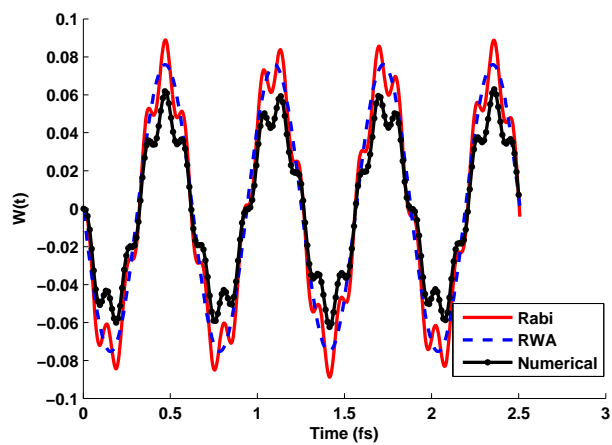}
\caption{Rabi oscillations of population inversion with a large detuning $\Delta=0.3\omega$.}\label{FIG6}
\end{figure}

Then, we consider the effect from the detuning factor $\Delta$. We fix $\Omega=0.02\omega$ and modify $\Delta=0.05\omega$ and then $\Delta=0.3\omega$. The results are shown in Fig. \ref{FIG5} and Fig. \ref{FIG6}. For a small detuning, there is a slight deviation between theoretical models and our numerical results. As the detuning becomes large, the discrepancy becomes more obvious. The Schr\"{o}dinger equation with the dipole approximation and ``length'' gauge [Eqs. \eqref{EQ44}--\eqref{EQ49}] breaks down in this situation.

\subsubsection{Effect from Loss}
\begin{figure}
\centering
\includegraphics[width=8cm]{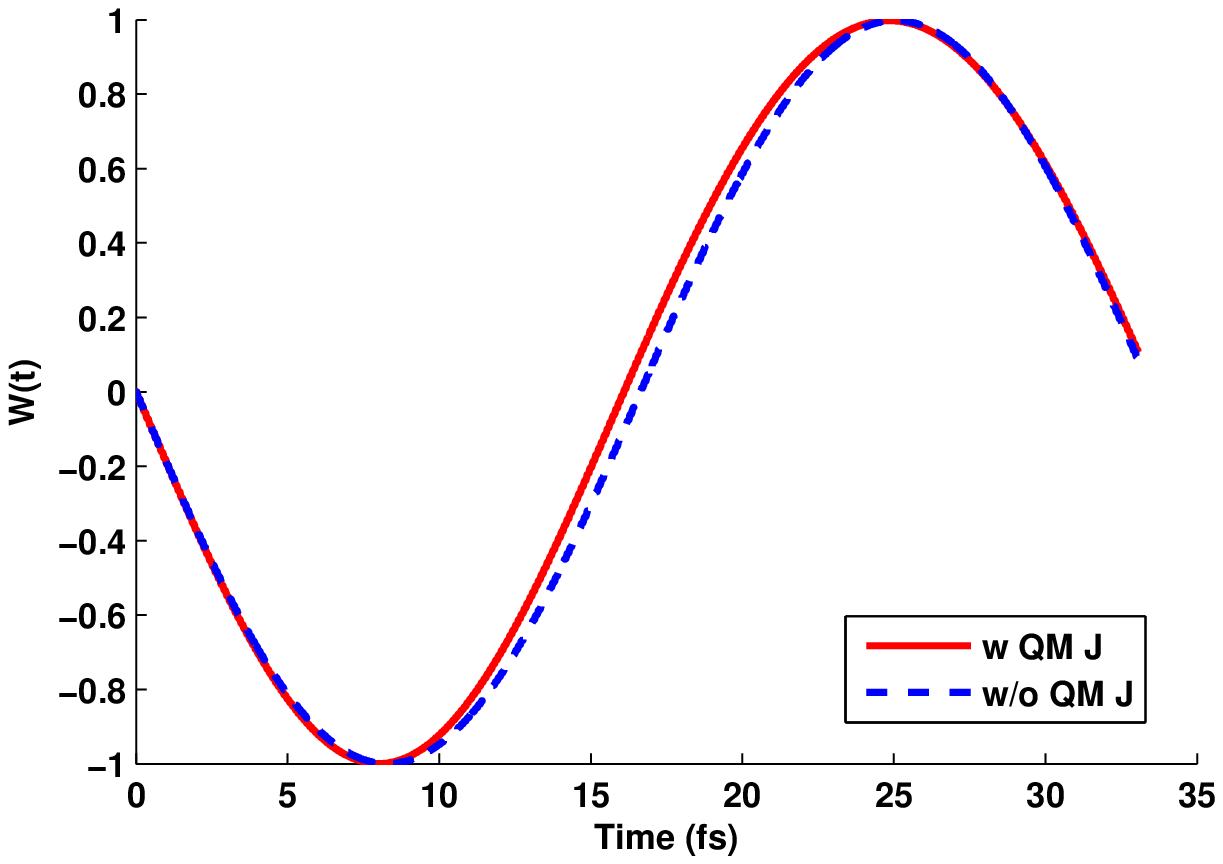}
\caption{Rabi oscillations of population inversion with and without the back coupling of the QM current for a small loss case.}\label{FIG7}
\end{figure}

\begin{figure}
\centering
\includegraphics[width=8cm]{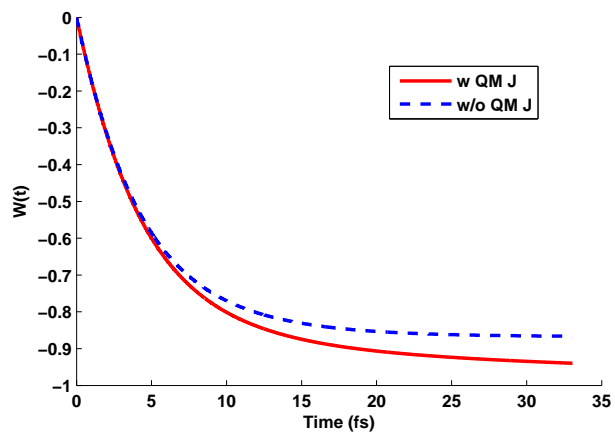}
\caption{Rabi oscillations of population inversion with and without the back coupling of the QM current for a large loss case.}\label{FIG8}
\end{figure}

Next, we investigate the effect from ohmic loss via filling conductive lossy meterial into the cavity. Since loss is introduced into our system, the EM fields will be attenuated and therefore the theoretical Rabi models cannot be adopted. The ohmic loss can be introduced via adding another conduction current $\mathbf{J}_c=\sigma \mathbf{E}$ to Eq. \eqref{EQ12}. Under the configuration of $\Omega=0.006\omega$ and $\Delta=0$, the results with small loss of $\sigma=0.0001$ and large loss $\sigma=0.02$ are shown in Fig. \ref{FIG7} and Fig. \ref{FIG8}. When the loss is small, the Rabi oscillation can be maintained cyclic. However, as the loss becomes larger, the EM field decays quickly, and the periodic Rabi oscillation can no longer be maintained. To measure the effect of the back coupling of the QM current, the Maxwell's equations are updated with and without the re-injection of the QM current (but the fore coupling is always considered, as is done in most QM calculations, namely, the Schr\"{o}dinger equation is solved considering the EM illumination). The feedback of the QM current to the EM system is relatively small when the loss is small, and such feedback becomes more significant when the loss becomes larger. Since a real system is always lossy, the unified coupled solution proposed in this paper will be preferred in real applications.

\subsubsection{Effect from Inhomogeneous Electromagnetic Environment}
\begin{figure}
\centering
\includegraphics[width=8cm]{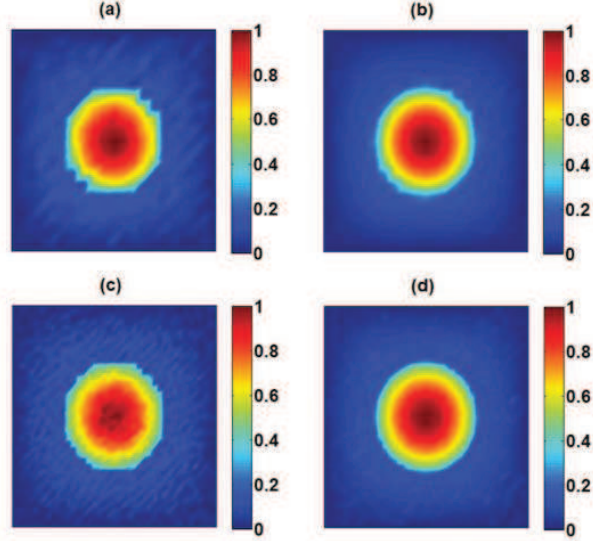}
\caption{{Fundamental mode of $Y_y$ component in an inhomogeneous cavity filled with a dielectric sphere. The observation plane is at the $xoz$ plane passing through the sphere center. (a) local-material approach with the spatial increment of 2 nm. (b) average-material approach with the spatial increment of 2 nm. (c) local-material approach with the spatial increment of 1 nm. (d) average-material approach with the spatial increment of 1 nm.}}\label{FIG8_add}
\end{figure}

{Finally, we investigate the influence of inhomogeneous electromagnetic environment on the population inversion of a particle. The particle is placed at the center of a dielectric sphere embedded in the cavity center. The cavity size is $40\,\,\mathrm{nm}\times 40\,\,\mathrm{nm} \times 40 \,\,\mathrm{nm}$ and the sphere has the radius of $10 \,\,\mathrm{nm}$ and relative permittivity of $4$. Except for the sphere, other regions of the cavity is filled with air. Here, the governing equation \eqref{EQ11} should be modified as}
\begin{equation}\label{EQ43_add}
\frac{{\partial {\bf{A}}}}{{\partial t}}  = \frac{\bf{Y}}{\epsilon_0\epsilon_r}
\end{equation}
{where $\epsilon_r$ is the position-dependent relative permittivity in the inhomogeneous electromagnetic system. $\mathbf{Y}=-\mathbf{D}$ and $\mathbf{D}$ is the electric flux. The fundamental mode of the system is calculated by the FDTD method. The local-material approach and the subcell based average-material scheme are adopted \cite{Sha2007} to treat the air-dielectric interface. As seen in Fig. \ref{FIG8_add}, the average-material scheme reduces the staircase error significantly; and the EM energy is confined at the sphere core.}

\begin{figure}
\centering
\includegraphics[width=8cm]{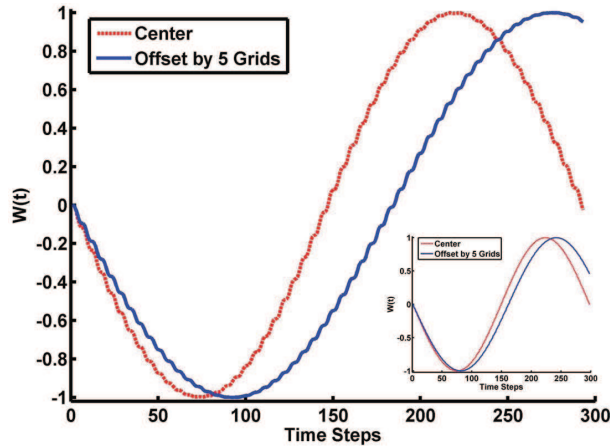}
\caption{{Population inversion of a particle in an inhomogeneous cavity filled with a dielectric sphere. The particle is located at the cavity center and that with an offset of 5 grids away from the center. The inset is given for the air-filled homogenous cavity with the same particle positions. The initial excitations are set to be the fundamental modes of the two cavities, respectively.}}\label{FIG8_add2}
\end{figure}

{To study the population inversion of the particle, the initial excitation is set to be the fundamental mode of the inhomogeneous cavity and the average-material scheme is employed. For comparisons, the results for the air-filled homogeneous cavity are also given and the initial excitation is set to be fundamental $\mathrm{TE}_{101}$ mode. For both cases, the particle is first located at the cavity center, where the EM excitation has the same maximum amplitude.  The transition frequencies of the particle are matched to the fundamental eigenfrequencies of the two cavities calculated by the FDTD method, respectively. Additionally, the particle position with an offset by 5 grids from the center is also considered. Due to a faster decay of the EM field away from the sphere center, the population inversion shows more significant lower Rabi frequency at the offset point for the inhomogeneous system (See Fig. \ref{FIG8_add2}).}

\subsection{Particle in free space}
\begin{figure}
\centering
\includegraphics[width=8cm]{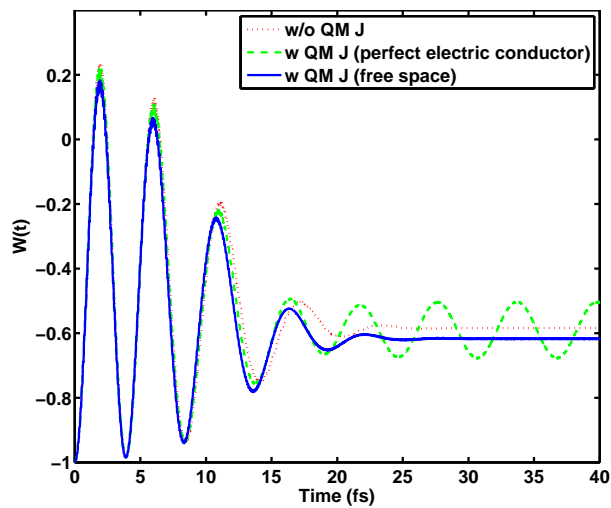}
\caption{Population inversion of a particle in free space. The self-consistent solution (with the back coupling of the QM current) and the non-self-consistent solution (without the back coupling of the QM current) are given for comparison. For the self-consistent solutions, results by the CPML absorbing boundary condition (free space) and perfect electric conductor boundary condition are also given for comparisons.}\label{FIG9}
\end{figure}

\begin{figure}
\centering
\includegraphics[width=8cm]{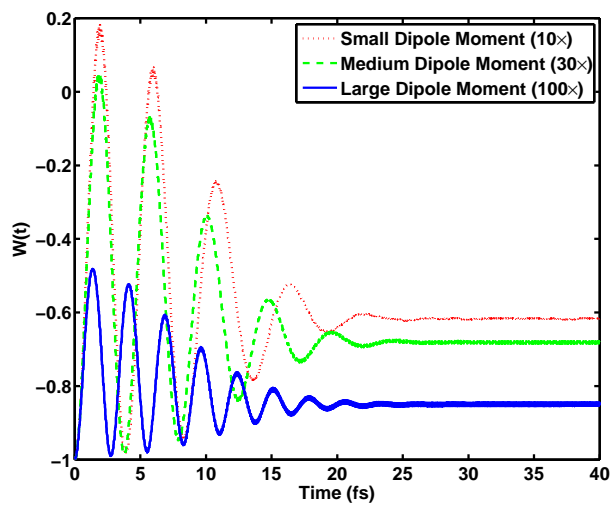}
\caption{Dipole moment dependent population inversion of a particle in free space. The self-consistent solution (with the back coupling of the QM current) is used.}\label{FIG10}
\end{figure}

{{A particle in the open free space is also investigated in this section. The computational domain for EM simulation occupies $40\times 40\times 40$ grids with a space increment of $1$ nm and a time increment of $6.75\times 10^{-4}$ fs. The total time steps are $6\times10^5$ with the CPU time around $30$ minutes. Ten CPMLs are employed to absorb radiated waves from the particle. The initial status of the particle is at its ground state with $a=1$ and $b=0$. The particle is driven by a predefined external field $\mathbf{E}(t)$ and the induced field $\mathbf{E}_{ind}(t)$ generated by the QM current. The external E-field is a cosine-modulated Gaussian pulse}}
\begin{equation}\label{eq45}
E_y(t)=   E_0 \cos \left( {\omega t} \right)\exp \left( { - \frac{{4\pi {{\left( {t - {t_0}} \right)}^2}}}{{{\tau^2}}}} \right)
\end{equation}
{{with an effective frequency range $[\omega/(2\pi)-2/\tau,\omega/(2\pi)+2/\tau]$ and $t_0=9\pi/(2\omega)$. Here, we set $E_0=10^{10}$ V/m, $\omega=33.3\times 10^{15}$ radians per second, $\tau=37.74$ fs and $t_0=0.42$ fs. The transition frequency is detuned from the EM frequency, i.e. $\omega_0=1.03\omega$. The external A-field with the Coulomb gauge can be obtained by a direct integration of the E-field, which is given by}}
\begin{equation}\label{eq46}
A_y(t)= -\int_0^t E_y(t) dt
\end{equation}
{{Considering the radiative dephasing (including the radiative decay and shift) is a very slow process, the dipole moment of the particle is set to be $10$ times of the quantum harmonic oscillator for observing the significant change of population inversion. In real applications, one can select molecules with the large dipole moment.}}

{{Figure \ref{FIG9} shows the calculated population inversion by the self-consistent solution (with the back coupling of the QM current) and by the non-self-consistent solution (without the back coupling of the QM current). For the self-consistent solutions, the results by the CPML absorbing boundary condition and the perfect electric conductor (PEC) boundary condition are also given for comparisons. After the external pulse is decayed to a negligible value, the Rabi oscillation of population is clearly observed for the PEC case. The radiated wave from the particle is reflected back by the PEC walls and is reabsorbed by the particle. Differently, the self-consistent solution of the population in free space (CPML case) keeps a constant and has a lower value compared to the non-self-consistent solution due to the radiative decay. Figure \ref{FIG10} depicts the dipole moment dependent population of the particle in free space. The self-consistent solution is adopted. A remarkable radiative decay and shift are observed as the dipole moment increases. As a result, the particle moves to its ground state.}}

\section{Conclusion}
A novel unified Hamiltonian approach is proposed to solve the coupled Maxwell-Schr\"{o}dinger equation self-consistently. This unified coupled system holds the energy conservation property during its time evolution. To overcome the multiscale issue caused by the distinct wavelength mismatch between the EM wave and electron wave, the reduced eigenmode expansion technique is adopted in the quantum system; and the relevant partial differential equations is cast into ODEs. Represented by the vector potential with a Coulomb gauge, Maxwell's equations are updated with the incorporation of quantum current obtained from the ODEs. Several physical settings (including field intensity, detuning, and material loss) affecting the Rabi oscillation are investigated and discussed, with comparison to theoretical approximate models. {{Furthermore, radiative decay and shift are also studied for the particle in free space. In future work, we will consider the EM field-particle interaction in more complex environment where material and structure dependent radiation and ohmic losses will modify the transition dynamics significantly.}}

\appendix
\section{Rabi Model and Rotating Wave Approximation}
Under dipole approximation and ``length'' gauge, the Schr\"{o}dinger equation in an EM environment can be expressed as \cite{Gerry2005}
\begin{equation} \label{EQ44}
\left[\frac{\hat{\mathbf{p}}^2}{2m} + V(\mathbf{r})-q\mathbf{r}\cdot\mathbf{E}\right]\Psi(\mathbf{r},t)
=i\hbar\frac{\partial\Psi(\mathbf{r},t)}{\partial t}.
\end{equation}
where the EM field is assumed to be monochromatic and is assumed not to be perturbed by the quantum system (namely no back coupling from the QM system)
\begin{equation} \label{EQ45}
\mathbf{E}=\mathbf{E}_0\cos(\omega t).
\end{equation}
Assuming a two-level system and applying reduced eigenmode expansion, we obtain a coupled set of ordinary differential equations:
\begin{equation}\label{EQ46}
i\hbar\frac{d{a(t)}}{d{t}}=-q\mathbf{E}_0\cdot\langle\psi_g|\mathbf{r}|\psi_e\rangle b(t) \cos(\omega t) e^{-i\omega_0t}
\end{equation}
\begin{equation}\label{EQ47}
i\hbar\frac{d{b(t)}}{d{t}}=-q\mathbf{E}_0\cdot\langle\psi_e|\mathbf{r}|\psi_g\rangle a(t) \cos(\omega t) e^{i\omega_0t}
\end{equation}
If we expand $\cos\left(\omega t\right)$ in exponentials and drop the term with fast oscillation frequency $\omega+\omega_0$, we have
\begin{equation}\label{EQ48}
i\hbar\frac{d{a(t)}}{d{t}}=-\frac{q\mathbf{E}_0}{2}\cdot\langle\psi_g|\mathbf{r}|\psi_e\rangle b(t) e^{i\left(\omega-\omega_0\right)t}
\end{equation}
\begin{equation}\label{EQ49}
i\hbar\frac{d{b(t)}}{d{t}}=-\frac{q\mathbf{E}_0}{2}\cdot\langle\psi_e|\mathbf{r}|\psi_g\rangle a(t) e^{-i\left(\omega-\omega_0\right)t}
\end{equation}
This is called the rotating wave approximation (RWA) and the equations can then be solved analytically if the atom is initially prepared at the ground state \cite{Gerry2005}:
\begin{equation} \label{EQ50}
a(t) = e^{i\Delta t/2}\left[\cos(\Omega_Rt/2)-i\frac{\Delta}{\Omega_R}\sin(\Omega_Rt/2)\right]
\end{equation}
\begin{equation} \label{EQ51}
b(t) = i\frac{\Omega}{\Omega_R}e^{i\Delta t/2}\sin(\Omega_Rt/2)
\end{equation}
where $\Delta=\omega-\omega_0$ is the detuning factor that measures the deviation of the atom transition frequency with the EM frequency, and $\Omega_R$ is the Rabi frequency
\begin{equation} \label{EQ53}
\Omega_R=\left(\Delta^2+\Omega^2\right)^\frac{1}{2}
\end{equation}
where
\begin{equation} \label{EQ52}
\Omega=\frac{-q\mathbf{E}_0\cdot\langle\psi_e|\mathbf{r}|\psi_g\rangle}{\hbar}.
\end{equation}


\section*{Acknowledgments}
This work was supported in part by NSFC 61201002, 61201122, in part by HK GRF 711511, UGC AoE/P¨C04/08, and in part by US AOARD 124082, 134140.


\end{document}